BRIEF COMMUNICATION

*25 March 2020 (revised on 21 May 2020 and 30 June 2020)*

# Distant sequence similarity between hepcidin and the novel coronavirus spike glycoprotein: a potential hint at the possibility of local iron dysregulation in COVID-19

Sepehr Ehsani [1,2,*]

[1] Theoretical and Philosophical Biology, Department of Philosophy, University College London, Bloomsbury, London, WC1E 6BT, United Kingdom
[2] Ronin Institute for Independent Scholarship, Montclair, New Jersey, 07043, United States

* E-mail: ehsani@uclmail.net / ehsani@csail.mit.edu

**ABSTRACT**

The spike glycoprotein of the SARS-CoV-2 virus, which causes COVID-19, has attracted attention for its vaccine potential and binding capacity to host cell surface receptors. Much of this research focus has centered on the ectodomain of the spike protein. The ectodomain is anchored to a transmembrane region, followed by a cytoplasmic tail. Here we report a distant sequence similarity between the cysteine-rich cytoplasmic tail of the coronavirus spike protein and the hepcidin protein that is found in humans and other vertebrates. Hepcidin is thought to be the key regulator of iron metabolism in humans. An implication of this preliminary observation is to suggest a potential route of investigation in the coronavirus research field making use of an already-established literature on the interplay of local and systemic iron regulation, cytokine-mediated inflammatory processes, respiratory infections and the hepcidin protein. The question of possible homology and an evolutionary connection between the viral spike protein and hepcidin is not assessed in this report, but some scenarios for its study are discussed.

**MAIN TEXT**

**1. Introduction**

As of the end of June 2020, 188 countries and regions are tackling the challenge of the pandemic caused by the novel coronavirus, with more than 10 million confirmed cases of infection worldwide (Center for Systems Science and Engineering, 2020). Coronaviruses, first described in the 1960s (Almeida et al., 1968; McIntosh, 2004), are mostly present in birds and mammals, and there has thus far been seven known coronavirus infectious disease outbreaks in humans causing respiratory illness (Hulswit et al., 2019; Seah et al., 2020). The four strains causing mild or common-cold-like symptoms are called 229E, NL63, OC43 and HKU1. The first two strains are in the 'alpha' coronavirus subgroup, whereas the latter two are 'beta' coronaviruses. The severe acute respiratory syndrome coronavirus (SARS-CoV-1) of 2002, the Middle East respiratory syndrome coronavirus (MERS-CoV) of 2012, and now the SARS-CoV-2 of 2019 (causing the 'COVID-19' disease) are the remaining three known coronaviruses (all of the beta subgroup) causing severe human disease (Andersen et al., 2020). This positive-sense single-stranded RNA virus family possesses the structural proteins spike (S), membrane (M) and envelope (E) proteins, along with the nucleocapsid (N) protein. It also has the largest genome among RNA viruses (Li, 2016).

Much research interest is devoted to the spike (glyco)protein (forming the characteristic 'corona') and its importance in the development of vaccines and antivirals (Beniac et al., 2006; Du et al., 2009). The angiotensin-converting enzyme 2 (ACE2) is thought to be its (main, but perhaps not exclusive) receptor on human host cells (Wrapp et al., 2020; Xia et al., 2020). The spike protein is formed of a receptor-binding subunit (S1), a membrane-fusion subunit (S2), a single-pass transmembrane (TM) domain, and a cytoplasmic/intracellular tail (CT) (Li, 2016; Shang et al., 2020). Of note, the S1 domain has a similar fold as human galectins (galactose-binding lectins) (Peng et al., 2011). Briefly, in terms of the putative *primary* function of the spike protein, Li comments that "because coronaviruses must enter cells for replication, membrane fusion is the central function of coronavirus spikes" (Li, 2016).

**2. Search for Sequence Similarity**

A basic question that might arise is: what exactly makes the pathobiology and disease course of these particular viruses *unique*? And, could it be that, in addition to viral replication inside the particular types of human host cells, other intracellular processes *specific* to these viruses are involved (for a thematically related inquiry, see e.g., (Gussow et al., 2020))? Having this in mind, we wondered if there might be any sequence similarity (and thereby potentially structural similarity) between the SARS-CoV-2 spike protein (which has 1,273 amino acids (A. Wu et al., 2020)) and any vertebrate protein(s).

A simple BLAST search does not reveal any similarities with human proteins. However, based on previous experience with the pufferfish *Takifugu rubripes* proteome (Ehsani, 2015; Ehsani et al., 2011; Schmitt-Ulms et al., 2009) and its unique evolutionary history, we initially restricted the search to this species. Teleost ancestor species underwent an ancient whole-genome duplication event (Van de Peer, 2004), and teleost species with well-annotated genomes such as the pufferfish can provide invaluable insights into the sequence evolution of genes with a clear phylogenetic linkage to mammalian genes (for a related commentary, see (Lidgard & Love, 2018)). Interestingly, a query using the full-length SARS-CoV-2 spike protein (accession no. YP_009724390.1) revealed a sole hit with the pufferfish hepcidin (Go et al., 2019) protein (XP_003965681.1; score: 32.7, E-value: 0.54). Given that SARS-like coronaviruses can be found in bats (F. Wu et al., 2020), we also used a full-length bat coronavirus sequence (ANA96027.1) as the query, which showed a closer match with pufferfish hepcidin (score: 38.5, E-value: 0.005). Conversely, a BLAST search in the *Coronaviridae* family of viruses using the pufferfish hepcidin (XP_003965681.1) revealed the bat coronavirus spike protein (ANA96027.1) as the top hit (score: 38.5, E-value: 5e-04). The scores and E-values here are not meant to indicate any claims of statistical significance but rather are provided for the purpose of comparison.

This similarity between the spike protein and hepcidin is at the cytoplasmic tail (McBride et al., 2007) of the spike protein, or, depending on the exact domain delineation, perhaps at the junction between the TM and CT domains. A multiple sequence alignment of this sequence region (generated using the AlignX feature of Vector NTI Advance 11.0, Invitrogen, Carlsbad, CA, USA), using three coronavirus spike proteins and four hepcidin proteins (from pufferfish, bat and human) is illustrated in **Figure 1A**.

The alignment depicts a number of conserved motifs, particularly between the first pufferfish hepcidin sequence (various pufferfish species have at least two hepcidin-like genes (Go et al., 2019)) and the coronavirus spike proteins. In a sense, the pufferfish sequence seems to act as a 'bridge' between the coronavirus motifs and those in the human hepcidin sequence. This may not be surprising particularly given the evolutionary context of teleost species described earlier. The similar cysteine-rich motif takes the following form: '**L**XXX**T**X**CC**X**CCKG**XXX**CG**X**CC**(**R/K**)**F**'. Of note, the eight cysteines of the mature human hepcidin in the similarity motif, and the aligned cysteines of the SARS-CoV-2 spike protein, are not all specifically coded by one of the two cysteine-coding codons (TGC and TGT). Both codons are present in the respective gene segments. Also, for comparison purposes, as the coronavirus envelope protein (Schoeman & Fielding, 2019) contains a related 'LCAYCCN' motif (Wu et al., 2003), this sequence was also added as the last line of the alignment.

Although this is a 'distant' and limited sequence similarity, it cannot be attributed to 'chance'. The search that found hepcidin did not reveal a *range* of similarities with other teleost proteins. Moreover, there are many cysteine-rich protein sequences in teleosts and vertebrates in general, yet this similarity to the hepcidin *gene family* (not merely a one-off sequence) was unique and specific. How or why this similarity arose is then a separate, and potential follow-up, question. On this point, rather than framing the question as one having to do with a chance/random occurrence, it could more aptly be framed under concepts related to convergent/adaptive evolution versus common ancestry. Clearly, however, no conclusive claims about homology and sequence conservation can be made at present without a concerted investigation focused on this topic. Nevertheless, the similarity reported here raises a potential and intriguing question of whether there could be mimicry of human hepcidin (structural, functional or otherwise), perhaps inside the host cell, by the TM-CT junction of the spike protein. These possibilities should not be dismissed based on the mere phylogenetic gap between coronaviruses and vertebrate species. Furthermore, the phylogenetic gap, and the absence of clear evolutionary homologies between genes that may otherwise have unexplored functional linkages, should not dissuade one from pursuing such connections. As a case in point, gene network and protein structure/function linkages between certain yeast (unicellular eukaryotes) and mammalian genes proved to be quite beneficial in investigating neurotoxicity in human cells (Tardiff et al., 2014). The question now is if the linkage between the spike protein and hepcidin could be expanded, and what reasonable scenarios for its experimental validation could be envisaged.

**3. The Hepcidin Protein**

Hepcidin is a small peptide hormone that was discovered in 2000/2001 (Andrews, 2008; Hunter et al., 2002; Krause et al., 2000; Park et al., 2001; Pigeon et al., 2001), and initially named LEAP-1 (liver-expressed antimicrobial peptide). It has an antiparallel beta-sheet fold and contains four disulfide bonds, and is involved in iron trafficking and the host's response to infection (Drakesmith & Prentice, 2012). In fact, it has been remarked by a number of commentators that "hepcidin is to iron, what insulin is to glucose" (Grover, 2019). Clear hepcidin orthologs appear to be missing in birds and invertebrates (Prentice, 2017). The human hepcidin (encoded by the *HAMP* gene on chromosome 19q13) is an 84-amino-acid prepropeptide, leading to a mature 25-amino-acid peptide that is detectable in blood and urine (Jordan et al., 2009). The proprotein convertase furin has been demonstrated to cleave prohepcidin at a polybasic site (Lee, 2008; Valore & Ganz, 2008). Of note, a furin-like cleavage site ('RRAR') has recently been reported to exist in the ectodomain of the SARS-CoV-2 spike protein, which is absent in coronaviruses of the same clade (Anand et al., 2020; Coutard et al., 2020). The protein-coding part of the *HAMP* gene is split over three exons, with the 25-amino-acid mature peptide occurring on the last exon. In terms of its putative function(s), Prentice notes that "although the hepcidin molecule does itself possess some antimicrobial activity, this is rather weak compared to peptides such as defensins, and its primary contribution to innate immunity is via regulation of iron" (Prentice, 2017). In fish species, one of the two hepcidin paralogs has been shown to potentially possess antimicrobial effects in innate immunity (B. Liu et al., 2020). Finally, hepcidin binds to and mediates the degradation of ferroportin (encoded by the *SLC40A1 / FPN1* gene), the only known cellular iron exporter. The structural details of this interaction are being mapped and studied in ever more detail (Billesbølle et al., 2020; Muriuki et al., 2019; Ren et al., 2020).

**4. Hepcidin and the CoV Spike Protein**

*4.1. Available Structures and Structural Predictions*

There are a number of solved structures of hepcidin (Jordan et al., 2009; Lauth et al., 2005). An NMR structure of human hepcidin is depicted in **Figure 1B** (visualized using 3-D Molecule Viewer, Vector NTI Advance 11.0, Invitrogen) including the locations of the four putative disulfide bonds. Of note, in addition to various hepcidin orthologs containing eight cysteines, four-cysteine variants have been described in notothenioid fish species (Xu et al., 2008). Available solved structures of the coronavirus spike glycoprotein, as far as our search could reveal, mostly utilize expression constructs that stop just short of the TM domain. As noted, this is partly because the protein's ectodomain is the main focus of studies on viral binding to host surface receptors (Walls et al., 2020; Wrapp & McLellan, 2019; Wrapp et al., 2020). For example, Wrapp, Wang and colleagues have reported the cryo-electron microscopy structure of ectodomain residues 1-1,208 of the spike protein (trimer in the prefusion conformation) (Wrapp et al., 2020), but this excludes the TM and CT domains. It goes without saying that the inclusion of transmembrane domains would require complicated structural elucidation protocols, and even then, one may still not be able to solve the structure of the protein in its entirety.

Moreover, we would like to note that using the Pfam-A (ver. 32) structural/domain database (Finn et al., 2008) in the HHpred remote homology and structure prediction toolkit (Zimmermann et al., 2018), the coronavirus spike protein regions analyzed here show some predicted structural similarity to lipolysis-stimulated receptor (LSR) lipoprotein receptor family (PF05624) (Yen et al., 1999), and hepcidin sequences show some predicted structural similarity to the Sar8.2 protein family found in Solanaceae plants (PF03058) (Alexander et al., 1992). What, if any, significance these findings may hold is unclear at present. Of more importance right now would be the theoretical and/or actual elucidation of the structure of the spike protein TM-CT junction region and a comparison with the available hepcidin structures (the Pfam hepcidin entry, PF06446, currently references six PDB structures). Of note, a recent *in silico* study has reported on a predicted structural similarity and compatibility between hepcidin and an allosteric site in the SARS-CoV-2 spike protein (Di Paola et al., 2020).

*4.2. Posttranslational Modifications*

Given the prominence of cysteines in the aligned motif (**Figure 1A**), how are they utilized in the respective similar domains? At first pass, the usages appear to be different: as noted earlier, in hepcidin, the cysteines may give rise to a compact disulfide-bridged peptide (Jordan et al., 2009; Nemeth & Ganz, 2009) (**Figure 1B**), whereas in coronavirus spike glycoproteins, the cysteines in the TM-CT junction serve as palmitoylation acceptor residues (Shulla & Gallagher, 2009) (**Figure 1C**) that facilitate membrane fusion (Chang et al., 2000). It should be mentioned, however, that 'mini-hepcidins' conjugated to palmitoyl groups have been synthesized and studied previously (Preza et al., 2011), but these do not occur naturally. As for the SARS-CoV, at least a portion of the palmitoylation of the spike protein has been reported to occur in a pre-medial Golgi compartment (McBride & Machamer, 2010). However, there is also the possibility of cross-disulfide bond formation with a non-homologous small cluster of cysteines within the envelope protein (Wu et al., 2003). Moreover, if *S*-palmitoylation is a reversible and dynamic process (Sobocinska et al., 2017), it is to be determined if the spike protein junction cysteines might in fact have a different posttranslational modification in the host cytoplasmic environment (although there is no evidence of this at the moment, and the reversibility of palmitoylation in viral spike proteins has not been reported (Gadalla & Veit, 2020)). That being said, in the cited paper by McBride and Machamer, the authors conclude that the palmitoylation of the SARS-CoV spike protein "was not necessary for S protein stability, trafficking or subcellular localization" nor "for efficient interaction with M protein" (McBride & Machamer, 2010). To what extent the posttranslational modifications of the spike protein and hepcidin, be it furin cleavage, disulfide bonds or palmitoylation, are in any way similar in an intracellular context, remains to be examined.

*4.3. Scenarios for the Investigation of Homology and Evolutionary Connections*

As alluded to in **Section 2**, in terms of the possibility of an evolutionary connection between the spike protein and hepcidin, one could imagine a scenario whereby an ancestral spike protein acquired a hepcidin-like sequence from a host organism, and the new sequence was palmitoylated to aid with membrane association. Li points out that "the primordial form of coronavirus spikes might contain S2 only" (Li, 2016), and the cytoplasmic motif features highlighted in the current report do not appear to be present in other class I viral membrane fusion proteins (which include the influenza virus (Bosch et al., 2003; Kielian, 2014)), although we have not performed an exhaustive search. However, a number of questions might then arise under this scenario, such as the difference between the primary localizations of hepcidin (considering its putative interaction with extracellular and transmembrane regions of ferroportin (Billesbølle et al., 2020; Ren et al., 2020)) versus the CT domain of the spike protein. Alternatively, it might be argued that perhaps a case of convergent sequence evolution is at play. For example, the influenza virus hemagglutinin glycoprotein appears to have a conserved 'CXICI' motif in its cytoplasmic tail domain (Nobusawa et al., 1991; Wagner et al., 2005), and perhaps an ancestral spike protein with similar features convergently acquired hepcidin-like sequence motifs. These are of course speculations and remain open questions. Investigations pursuing these topics could also make use of studies that attempt to trace the evolutionary history of hepcidin itself (Kim et al., 2019; Xu et al., 2008).

**5. Potential Leads for Coronavirus Research**

One of the first questions that may arise from this potential sequence similarity is whether the comparison is equal for all seven known human coronavirus strains, or if it is more limited to the severe disease-causing viruses. **Figure 2** depicts a sequence alignment to start to answer this question. The alignment is restricted to the length of the mature human hepcidin protein, which is depicted at the bottom row. The first four spike protein sequences are comprised of the mild/asymptomatic strains, i.e., 229E and NL63 (alpha coronaviruses) and HKU1 and OC43 (beta coronaviruses). These are followed by the spike proteins of MERS-CoV, SARS-CoV-1 and SARS-CoV-2. The bat coronavirus sequence and the four hepcidins are depicted in the same order as in **Figure 1A**. Paying particular attention to the region between the two conserved cysteines (marked by two black arrows), there appears to be less similarity in the motif between the first four spike proteins than the rest of the sequences. Specifically, for example, a 'conserved' glycine in the sixth position from the first overall-conserved cysteine (indicated with a red arrow) appears to be an important residue that groups together the three disease-causing spike proteins with the bat sequence and the two teleost hepcidin proteins. How, if at all, these differences play out in the cell might be an intriguing experimental direction.

What can also be noted from **Figure 2** is that although there are appreciable differences in the spike protein domains visualized in the figure between the MERS-CoV and the two SARS-CoV sequences, the SARS-CoV-1 and SARS-CoV-2 spike proteins are almost identical in this region (the tabulated amino acid identity and conservation values shown were calculated for the aligned region using the AlignX program). Therefore, based on the hypothetical link presented in this report, it might be reasonable to assume that any pathobiological differences between the two SARS strains would not be as a result of any differing biology attributable to the hepcidin similarity domain. For reference, the SARS-CoV-2 genome has been reported to be 96% identical to bat CoV (nucleotide identity), 80% to SARS-CoV-1, 55% to MERS-CoV and 50% to common cold CoV (e.g., the 229E and OC43 strains) (Bar-On et al., 2020). Also of note in **Figure 2** is the percentage residue identity and conservation between *Takifugu* hepcidin and the corresponding bat coronavirus spike protein region (54% identity, 62% conservation), which are the same identity and conservation values between the two *Takifugu* hepcidin paralogs.

More broadly, the potential sequence link reported here points to the need to build on previous research on the cytoplasmic tail of the spike protein (e.g., (McBride et al., 2007; Petit et al., 2007; Petit et al., 2005)) to better understand its distinct roles from the time of viral attachment to possible protein-protein interactions inside the host cell. Of the rare mutations currently reported in the SARS-CoV-2 spike protein (using patient-isolated genomic data in the GISAID repository), only one (P1263L) appears to be in the cytoplasmic tail domain (Korber et al., 2020), placing it outside of the similarity region reported here.

The sequence similarity also points to a focus on iron biology. Following calcium, oxygen and lead, iron has historically been one of the most studied elements in cell biology (Ehsani, 2013), and as such there is a vast body of (at times conflicting) literature to draw upon relating to iron in coronavirus infections. Other investigators have reviewed broad themes from iron biology in the context of COVID-19 (Edeas et al., 2020; W. Liu et al., 2020), and so here I will only stay on the topic of hepcidin. Foremost among the potential physiological connections related to this discussion is the so-called 'cytokine storm' mediated by interleukin-6 (IL-6) reported in some COVID-19 patients (Laing et al., 2020; Mehta et al., 2020; Moore & June, 2020; Xu et al., 2020). Indeed, researchers have recently proposed that "reduced innate antiviral defenses coupled with exuberant inflammatory cytokine production are the defining and driving features of COVID-19" (Blanco-Melo et al., 2020). Hepcidin production in the liver is induced by IL-6, and this is especially well-studied in the anemia of inflammation literature (Ganz, 2019; Noguchi-Sasaki et al., 2016). Related to the role of IL-6, it has been reported that hepatic heparan sulfate affects and regulates IL-6-stimulated hepcidin expression (Poli et al., 2019). Furthermore, heparin, the anticoagulant glycosaminoglycan that is a highly sulfated form of heparan sulfate, has been shown to be a potent inhibitor of hepcidin expression (Poli et al., 2011). Of interest, anticoagulant treatment has been reported to be effective in a subset of severe COVID-19 patients (Tang et al., 2020).

Less established, but still important to mention, is the current debate surrounding the issue of hypoxia/hypoxemia and certain symptoms resembling, but differentiable from, altitude illness (Archer et al., 2020; Arias-Reyes et al., 2020; Basnyat et al., 2020; Gattinoni et al., 2020). Congruously, hepcidin expression levels have also been investigated in the context of high-altitude acclimatization (Hennigar et al., 2020) (see also (Choque-Quispe et al., 2020)). Moreover, hepcidin levels are known to increase in patients with acute respiratory distress syndrome (ARDS) (Galesloot et al., 2011; Moccia et al., 2020). A second, not yet fully-established, link of relevance here is the recent observations of a Kawasaki-disease-like systemic vasculitis syndrome in children infected with the novel coronavirus (Jones et al., 2020; Verdoni et al., 2020), which is also being called multisystem inflammatory syndrome in children (MIS-C) (Dufort et al., 2020; Feldstein et al., 2020). Incidentally, an association between a novel human coronavirus and Kawasaki disease was reported in 2005 (Esper et al., 2005), although other investigators were apparently not successful in confirming the link (Ebihara et al., 2005). Nevertheless, if the association with the Kawasaki-disease-like syndrome is real, then it is noteworthy that increased hepcidin levels have been suggested as a biomarker for Kawasaki disease (Huang et al., 2018).

Given the central role of hepcidin in iron metabolism, it is also important to point to a number of circumstantial but perhaps important findings in the literature pertaining to lung disease. These include (i) a link between SARS and liver function abnormalities (Lefkowitch, 2005), (ii) the association of pulmonary iron overload and restrictive lung disease (Ganz, 2017; J. Neves et al., 2017), (iii) the role of iron in pulmonary fibrosis (Ali et al., 2020), and (iv) hepcidin's modulation of the proliferation of pulmonary artery smooth muscle cells (Ramakrishnan et al., 2018). Furthermore, (v) hepcidin upregulation along with serum iron reduction has been reported in influenza infections (Armitage et al., 2011; Fernandez, 1980). Importantly, however, iron dysregulation changes may only be at a local cellular/tissue level and not reach a systemic response (Lakhal-Littleton et al., 2019), and may demonstrate selective tissue tropisms during different viral infections (Armitage et al., 2014).

Moving back to the biology of the hepcidin protein itself, as noted previously, hepcidin binds to and mediates the degradation of the iron exporter ferroportin. If the sequence similarity reported here is actually playing a significant role at the cellular level, could it be that, although the cellular localizations appear to be different based on current knowledge, the SARS-CoV-2 spike protein cytoplasmic tail can partly mimic the structure of hepcidin and interact with ferroportin? Could the cytoplasmic tail even coordinate and bind iron? These remain to be investigated experimentally, but of note, Rishi and colleagues recently reported on the intracellular localization of ferroportin dimers, and concluded that both the carboxy- and amino-termini of the protein are intracellular (Rishi et al., 2020). As cited earlier, the details of hepcidin's own interaction sites with ferroportin remain the subject of different structural determination projects (Billesbølle et al., 2020; Ren et al., 2020). Relatedly, and of interest, Neves and colleagues, using experiments on iron overload in European bass (*Dicentrarchus labrax*), have discussed the functional partnership between hepcidin and ferroportin from an evolutionary perspective and suggested that this may "open new possibilities for the pharmaceutical use of selected fish […] hepcidins during infections, with no impact on iron homeostasis" (J. V. Neves et al., 2017).

Finally, and specific to COVID-19, a number of broader questions that could follow up from this work are: first, does the spike protein, similar to hepcidin, potentially promote iron sequestration in (alveolar) macrophages (Michels et al., 2015) and hence impede the host's immunological response? Second, could a recent report of the common presence of digestive symptoms in COVID-19 patients (Pan et al., 2020) be explainable in part by a link to hepcidin? And third, could systemic changes in serum iron levels (Hippchen et al., 2020; Shah et al., 2020; van Swelm et al., 2020; Zhang & Liu, 2020) (with a possible view on the degree of above-normal serum ferritin in patients (Chen et al., 2020)) or levels of hepcidin itself be detected in patients with varying COVID-19 severities? With respect to the biology surrounding reports of gastrointestinal symptoms, it is important to point out that a recent study concluded that in gut epithelial cells, the "expression of two mucosa-specific serine proteases, TMPRSS2 and TMPRSS4, facilitated SARS-CoV-2 spike fusogenic activity and promoted virus entry into host cells" (Zang et al., 2020). It is also known that the liver-specific serine protease TMPRSS6 (matriptase-2) negatively modulates hepcidin (Beliveau et al., 2019; Camaschella, 2015; Du et al., 2008). Overall, the observations in this report suggest that, as a starting point, (serum) iron status would be a critical data category to be *systematically* collected from patients at various stages of the disease's progression. Serum ferritin levels are usually considered a general indicator of inflammation and infection, and as pointed out by Baron and colleagues, "the use of ferritin to diagnose iron deficiency may be problematic in patients with COVID-19 disease, who may have normal or high ferritin levels despite very low iron stores" (Baron et al., 2020; Zhou et al., 2020). Be that as it may, iron homeostasis in the case of COVID-19 may be a more *specific* pathobiological feature of this infection, and we can only know the answer to this scenario if more data related to iron biology is gathered during the current pandemic.

## 6. Conclusion

Theoretical analyses in new areas may necessarily entail reasonable speculations based on limited or disparate data. This is expected, but one should remain cognizant of overinterpretation, and pursue a rational course of theoretical inquiry to hopefully inform subsequent experimental investigations. Here, a purposeful and restricted protein sequence search revealed a potential sequence similarity between the relatively less-studied cysteine-rich cytoplasmic domain of coronavirus spike proteins and the vertebrate hepcidin protein. This is quite unlikely to be a spurious and random similarity. There are many cysteine-rich protein sequences in vertebrates, but the motif identified here is unique and specific, and also appears to tentatively set apart the disease-causing strains from the milder coronavirus strains. Following from this link, a number of emerging clinical strands of evidence (summarized in **Table 1**) were discussed which further link a biology surrounding hepcidin with coronavirus-caused pathobiology. While each piece of clinical evidence discussed does not by itself provide overwhelming corroboration for the hypothesis of the paper, the totality of evidence presented we believe make a strong case that if sequence and/or structural mimicry to hepcidin is taking place upon viral attachment to and entry in the host cell, then perhaps a local disease condition resembling iron dysregulation (e.g., iron overload) might result in the infected tissue(s). This hypothesis can be immediately tested in one of three ways. First, in the clinic, levels of various serum markers for iron biology could be more systematically and comprehensively measured and analyzed. Second, computational investigations could examine potential structural mimicry between the two proteins and explore the effect of differing posttranslational modifications. And third, the potential link to hepcidin could be relied upon in cell-based assays to determine the possibility of the involvement of the spike protein in iron biology.

# TABLE

**Table 1.** Points of similarity and divergence between hepcidin and the coronavirus spike protein cytoplasmic domain based on current knowledge of the viral protein and its pathobiology (see main text for references)

| Potential Similarities | Potential Differences |
|---|---|
| 1. **Protein sequence:** A unique but restricted sequence similarity exists between mature hepcidin and the cysteine-rich cytoplasmic tail of coronavirus spike proteins. | 1. **Protein length and domains:** Coronavirus spike proteins are much larger, multi-domain, proteins compared to the relatively short-length hepcidin proteins. |
| 2. **Posttranslational processing:** The proprotein convertase furin cleaves prohepcidin, and has been reported to also activate (the ectodomain of) the SARS-CoV-2 spike protein. | 2. **Posttranslational modification:** Cysteines in the cytoplasmic tail of coronavirus spike proteins are thought to be palmitoylation acceptor residues, and this modification has thus far not been reported to be reversible. Cysteines in hepcidin are used to form disulfide bonds. |
| 3. **Cytokine storm:** IL-6-mediated inflammatory responses have been reported in COVID-19 patients. Hepcidin production in the liver is induced by IL-6 and is well-studied in the context of anemia of inflammation. In addition, COVID-19 may be associated with a Kawasaki-disease-like systemic vasculitis manifestation in children, and hepcidin levels have also been suggested as a biomarker for Kawasaki disease. | 3. **Localization:** Hepcidin is thought to interact with its main (iron-bound) interactor ferroportin extracellularly, whereas the spike protein cytoplasmic tail does not face the environment outside the viral membrane. (However, the cytoplasmic tail associates with the plasma membrane itself aided by its palmitoylation modifications, and hepcidin may interact with ferroportin close to ferroportin's transmembrane regions in addition to extracellular residues.) |
| 4. **Hypoxia/Hypoxemia:** COVID-19 may lead to symptoms resembling, but differentiable from, altitude illness, and hepcidin expression levels have also been studied in the context of high-altitude acclimatization. | |

**FIGURE LEGENDS**

**Figure 1. Comparison of select hepcidin and coronavirus spike protein sequences.** (**A**) A multiple sequence alignment of the C-terminal region of a number of coronavirus spike proteins (encompassing portions of the putative transmembrane and cytoplasmic tail segments), four hepcidins and the SARS-CoV-2 envelope protein is presented. The envelope sequence is provided only to demonstrate the cysteine residues with which the spike protein is proposed to form disulfide bridges (Wu et al., 2003). The residue numbers are shown on the sides of each protein segment, and for proteins whose C-terminal sequences continue beyond the alignment, the full residue length is provided to the right. As per a color scheme used previously (Schmitt-Ulms et al., 2009), dark green, grey and black highlights depict conserved, similar and identical residues, respectively. 'Tr' stands for *Takifugu rubripes* (Japanese pufferfish), 'Rf' for *Rhinolophus ferrumequinum* (greater horseshoe bat) and 'Hs' for *Homo sapiens* (human). The protein accession numbers of the sequences shown are, in order: (1) AWH65954.1, (2) YP_009724390.1, (3) ANA96027.1, (4) XP_003965681.1, (5) XP_029694670.1, (6) ENSRFET00010014064.1, (7) NP_066998.1 and (8) QHD43418.1. The domain illustration of the spike protein is based on (Wrapp et al., 2020). (**B**) A solved NMR structure of human hepcidin (Jordan et al., 2009) (PDB: 2KEF), adopting an antiparallel beta-sheet fold, is visualized with its putative four disulfide bonds formed between eight cysteine residues. (**C**) The position of the disulfide bonds in the sequence of the mature human hepcidin is illustrated along with the potential palmitoylation residues (ten cysteines) of the cytoplasmic tail of the SARS-CoV-2 spike protein. The palmitate visual is as per (Linder & Deschenes, 2007).

**Figure 2. Comparison of core hepcidin-like motif among the seven known coronavirus human infection-causing strains.** Following the alignment presented in **Figure 1A**, the core similarity motif, corresponding to the length of the mature human hepcidin sequence, is shown in an alignment containing the spike protein cytoplasmic domain of the four mild/asymptomatic human coronavirus strains and the three severe-disease-associated strains. Hepcidin sequences from the previous figure appear on the last four lines. The alignment color scheme is the same as in **Figure 1**. Focusing on the region between the two black arrows, fewer similar residues could be observed between the group of mild/asymptomatic coronavirus strains and the MERS/SARS-CoV strains. One such residue that can act to distinguish the groups is indicated with a red arrow. The accession numbers of the sequences shown, in order of appearance, are: (1) ARU07601.1, (2) AFD98827.1, (3) AZS52618.1, (4) AXX83351.1, (5) AWH65954.1, (6) NP_828851.1, (7) YP_009724390.1, (8) ANA96027.1, (9) XP_003965681.1, (10) XP_029694670.1, (11) ENSRFET00010014064.1 and (12) NP_066998.1. The table presenting the amino acid identity and conservation for the sequences in the aligned region shows the values in percentage points, with the residue conservation values appearing in brackets.

**ACKNOWLEDGEMENTS**

An earlier draft of this manuscript was posted on 27 March 2020 on the *arXiv* preprint server (arxiv.org/abs/2003.12191v1). The author declares no competing interests, and is supported by a University College London Overseas Research Scholarship. Thanks are due to Gerold Schmitt-Ulms (University of Toronto) and Farnaz Fahimi Hanzaee (University College London) for very helpful discussions.

# FIGURE 1

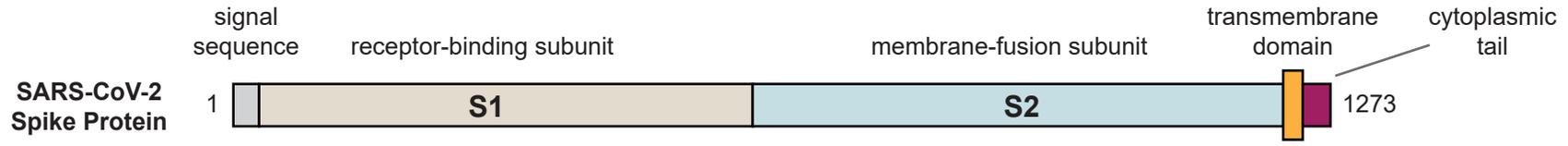

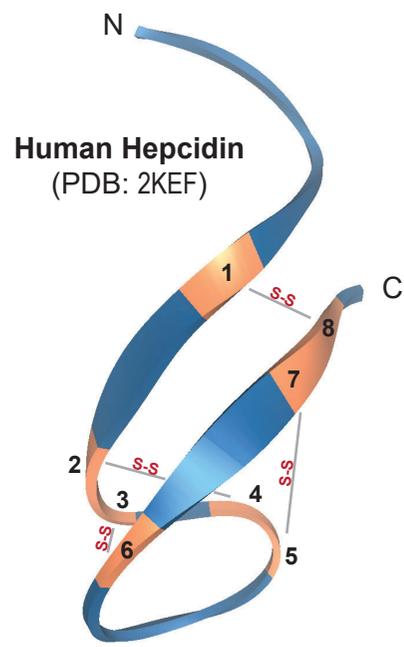

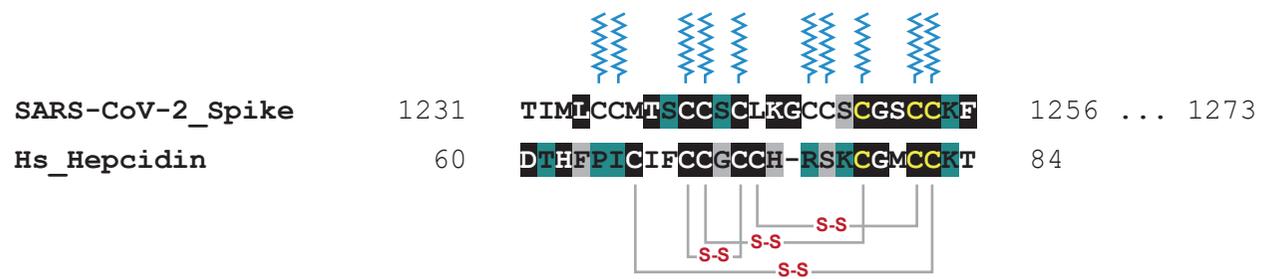

```
HCoV-229E_Spike    1131   LLLCCCSTGCCGFFSCFASSIKGCCE    1156 ... 1171
HCoV-NL63_Spike    1314   LVFCCLSTGCCGCCNCLTSSMRGCCD    1339 ... 1356
HCoV-HKU1_Spike    1313   LFFICCCTGCGSACF--SKCHNCCDE    1336 ... 1351
HCoV-OC43_Spike    1324   LFFICCCTGCGPSCF--KKCGGCCDD    1347 ... 1362
MERS-CoV_Spike     1308   VFFLLCCTGCGTNCMGKLKCNRCCDK    1333 ... 1347
SARS-CoV-1_Spike   1213   TILLCCMTSCCSCLKGACSCGSCCKF    1238 ... 1255
SARS-CoV-2_Spike   1231   TIMLCCMTSCCSCLKGCCSCGSCCKF    1256 ... 1273
Bat_CoV_Spike      1194   IILLCYFTSCCSCCKGMCSCGSCCRF    1219 ... 1236
Tr_Hepcidin-1      65     QSHLSLCTLCCNCCKGNKGCGFCCRF    90
Tr_Hepcidin-2      63     KRSPKRCKFCCNCCPGMRGCGVCCRF    88
Rf_Hepcidin        103    DAHFPICIYCCGCCY-KSRCGLCCKT    127
Hs_Hepcidin        60     DTHFPICIFCCGCCH-RSKCGMCCKT    84
```

Amino Acid Identity (and Conservation) (%)

| | HCoV-NL63_Spike | HCoV-HKU1_Spike | HCoV-OC43_Spike | MERS-CoV_Spike | SARS-CoV-1_Spike | SARS-CoV-2_Spike | Bat_CoV_Spike | Tr_Hepcidin-1 | Tr_Hepcidin-2 | Rf_Hepcidin | Hs_Hepcidin |
|---|---|---|---|---|---|---|---|---|---|---|---|
| HCoV-229E_Spike | 58 (77) | 33 (42) | 29 (33) | 19 (31) | 31 (42) | 27 (42) | 27 (42) | 15 (19) | 12 (15) | 16 (20) | 16 (20) |
| HCoV-NL63_Spike | | 33 (46) | 38 (38) | 27 (31) | 27 (38) | 27 (38) | 31 (50) | 27 (31) | 19 (27) | 24 (32) | 24 (36) |
| HCoV-HKU1_Spike | | | 75 (83) | 58 (79) | 33 (46) | 33 (46) | 33 (50) | 29 (38) | 25 (29) | 29 (42) | 33 (38) |
| HCoV-OC43_Spike | | | | 58 (71) | 33 (46) | 33 (46) | 33 (50) | 38 (42) | 29 (33) | 29 (38) | 33 (33) |
| MERS-CoV_Spike | | | | | 31 (38) | 31 (38) | 31 (42) | 35 (35) | 27 (27) | 28 (32) | 28 (32) |
| SARS-CoV-1_Spike | | | | | | 92 (96) | 77 (81) | 46 (58) | 35 (46) | 32 (36) | 32 (36) |
| SARS-CoV-2_Spike | | | | | | | 73 (81) | 46 (58) | 35 (46) | 32 (36) | 32 (36) |
| Bat_CoV_Spike | | | | | | | | **54 (62)** | 46 (54) | 32 (40) | 32 (40) |
| Tr_Hepcidin-1 | | | | | | | | | **54 (62)** | 40 (52) | 40 (52) |
| Tr_Hepcidin-2 | | | | | | | | | | 36 (48) | 40 (48) |
| Rf_Hepcidin | | | | | | | | | | | 76 (96) |